\documentclass{pasj00}
\draft

\newcommand{\suzaku}{{\it Suzaku}\ }
\newcommand{\chandra}{{\it Chandra}\ }
\newcommand{\rosat}{{\it ROSAT}\ }
\newcommand{\asca}{{\it ASCA}\ }
\newcommand{\astroh}{ASTRO-H\ }

\usepackage{graphicx,times}

\begin{document}
\SetRunningHead{Tamura et al.}{Gas Bulk Motion in of A~2256}
%\Received{2008/02/22}%{yyyy/mm/dd}
%\Accepted{2008/03/26}%{yyyy/mm/dd}

\title{Discovery of Gas Bulk Motion in the Galaxy Cluster Abell 2256 with {\it Suzaku} }

%%% begin:list of authors
\author{
Takayuki \textsc{Tamura}\altaffilmark{1},
Kiyoshi \textsc{Hayashida}\altaffilmark{2},
Shutaro \textsc{Ueda}\altaffilmark{2},
Masaaki \textsc{Nagai}\altaffilmark{2},
}
\altaffiltext{1}{
Institute of Space and Astronautical Science,
Japan Aerospace Exploration Agency,\\
3-1-1 Yoshinodai, Sagamihara, Kanagawa 229-8510
}
\email{tamura.takayuki@jaxa.jp}
\altaffiltext{2}{
Department of Earth and Space Science, Graduate School of Science, Osaka University, Toyonaka 560-0043
}


\maketitle

\begin{abstract}
The results from \suzaku observations of the galaxy cluster Abell~2256 are presented.
This cluster is a prototypical and well-studied merging system, 
exhibiting substructures both in the X-ray surface brightness 
and in the radial velocity distribution of member galaxies.
There are main and sub components 
separating by \timeform{3'.5} in the sky and by about 2000 km s$^{-1}$ in radial velocity peaks of member galaxies.
In order to measure Doppler shifts of iron K-shell lines from the two gas components by the \suzaku XIS,
the energy scale of the instrument was evaluated carefully and found to be calibrated well.
A significant shift of the radial velocity of the sub component gas with respect to that of the main cluster
was detected.
All three XIS sensors show the shift independently and consistently among the three.
The difference is found to be 1500 $\pm 300$ (statistical)  $\pm 300$ (systematic) km s$^{-1}$.
The X-ray determined absolute redshifts of and hence the difference between the main and sub components are consistent with those of member galaxies in optical.
The observation indicates robustly that the X-ray emitting gas is moving together with galaxies 
as a substructure within the cluster.
These results along with other X-ray observations of gas bulk motions in merging clusters
are discussed. 

\end{abstract}

\section{Introduction}
\label{sect:intro}
%\today\\
% 1. background, topics, cluster merger, dynamics\\
Galaxy clusters are the largest and youngest gravitationally bound system among the hierarchical structures in the universe.
Dynamical studies of cluster galaxies have revealed that some systems are still forming and unrelaxed.
X-ray observation of the intracluster medium (ICM) 
provided further evidence for mergers through spatial variations of the gas properties.
Remarkably, sharp X-ray images obtained by \chandra have revealed shocks (e.g. \cite{m02}) and density discontinuities (or ``cold fronts''; e.g. \cite{v01}).
These are interpreted as various stages of on-going or advanced mergers
and suggest supersonic (for the shock) or transonic (cold front) gas motions. 
Cluster mergers involve a large amount of energies
and hence influence numerous kinds of observations.
In particular, possible effects of gas bulk motions on the X-ray mass estimates
have been investigated extensively mostly based on numerical simulations (e.g. \cite{evrard96}, \cite{nagai07}).
This is mainly because that the cluster mass distribution is one of the most powerful tools for the precise cosmology.
Furthermore, cluster mergers heat the gas, develop gas turbulence, and accelerate particles, which in turn generate diffuse radio and X-ray halos.

% 2. importance, motivation for the gas bulk motion)\\
To understand physics of cluster mergers, 
gas dynamics in the system should be studied.
The gas bulk motion can be measured most directly using the Doppler shift of X-ray line emission.
These measurements are still challenging because of the limited energy resolutions of current X-ray instruments.
Dupke, Bregman and their colleagues searched for bulk motions using \asca in nearby bright clusters.
They claimed detections of large velocity gradients,
such as that consistent with a circular velocity of $4100^{+2200}_{-3100}$~km s$^{-1}$ (90\% confidence) in the Perseus cluster \citep{dupke01-per} and that of $1600 \pm 320$ ~km s$^{-1}$ in the Centaurus cluster\footnote{The error confidence range is not explicitly mentioned in the reference.} \citep{dupke01-cen}.
These rotations imply 
a large amount of kinetic energy comparable to the ICM thermal one.
Note that they used the \asca instruments (GIS and SIS) which have gain accuracies of about 1\% (or 3000~km s$^{-1}$).
\citet{dupke06} used also \chandra data and claimed a confirmation of the motion in the Centaurus cluster.
These important results, however, have not yet been confirmed by other groups. 
For example, \citet{ezawa01} used the same GIS data of the Perseus cluster 
and concluded no significant velocity gradient.
In addition, \citet{ota07} found that the \suzaku results of the Centaurus cluster 
are difficult to reconcile with claims in \citet{dupke01-cen} and \citet{dupke06}.
In short, previous results by Dupke and Bregman suggest bulk motions in some clusters but with large uncertainties.
% We attemped to improve the accuracy of the velocity determination using new instrument.

%3.
Currently the \suzaku XIS \citep{koyama07} would be the best X-ray spectrometer for the bulk motion search,
because of its good sensitivity and calibration \citep{ozawa09}.
In fact, \suzaku XIS data were already used for this search in representative clusters.
Tight upper limits on velocity variations are reported 
from the Centaurus cluster (1400 km s$^{-1}$; 90\% confidence; \cite{ota07}) 
and A~2319 (2000 km s$^{-1}$; 90\% confidence; \cite{sugawara09})
among others.

% 4. A2256 and Suzaku Observation, what is new, merit of instrument, target\\
In order to improve the accuracy of the velocity determination and to search for gas bulk motions 
we analyzed \suzaku XIS spectra of the Abell 2256 cluster of galaxies (A~2256, redshift of 0.058).
This X-ray bright cluster is one of the first systems showing substructures not only in the X-ray surface brightness 
but also in the galaxy velocity distribution \citep{briel91}.
In the cluster central region, there are two systems separating by \timeform{3'.5} in the sky.
Motivated by this double-peaked structure in their \rosat image, 
\citet{briel91} integrated the velocity distribution of galaxies from \citet{fkk89} over the cluster,
fitted it to two Gaussians, and found the two separated peaks in the velocity distribution.
The two structures are separated by $\sim 2000$ km s$^{-1}$ in radial velocity peaks of member galaxies, 
as given by table~\ref{tbl:velocity}. 
\citet{Berrington02} added new velocity data to the \citet{fkk89} sample, used 277 member galaxies in total, 
and confirmed the two systems along with an additional third component (table~\ref{tbl:velocity}).
This unique finding motivated subsequent observations in multiple wavelengths.
For example,  
radio observations revealed a diffuse halo, relics, and tailed radio emission from member galaxies (e.g. \cite{rot94}).
The \chandra observation by \citet{sun02} revealed detailed gas structures in and around the main and second peaks. 
Furthermore, there are some attempts to reproduce merger history of A~2256 using numerical simulations (e.g. \cite{roe95}).
Thus, A~2256 is a prototypical and well-studied merging system
and hence suitable to study the gas dynamics.

\begin{table}[h]
  \caption{Fitting parameters of radial velocity distribution of member galaxies.}
\label{tbl:velocity}
  \begin{center}
    \begin{tabular}{lll}
\hline \hline
%References (and component number)  & Mean velocity & Velocity $\sigma$  \\
component number & Mean velocity & Dispersion \\
in the reference & (km~s$^{-1}$) & (km~s$^{-1}$)\\
\hline
\multicolumn{3}{c}{\citet{briel91}} \\
1 (main) & $17880\pm 205$ & $1270\pm 127$ \\
2 (sub) & $15730\pm 158$ & $350\pm 123$ \\
\hline
\multicolumn{3}{c}{\citet{Berrington02}\footnotemark[$*$] } \\
1 (53 galaxies; sub) & 15700 & 550 \\
2 (196 galaxies; main) & 17700 & 840 \\
3 (28 galaxies) & 19700 & 300 \\
\hline
\multicolumn{3}{@{}l@{}}{\hbox to 0pt{\parbox{85mm}{\footnotesize
\par\noindent
\footnotemark[$*$] Errors on velocities and dispersions are not given in the paper.
}\hss}}
    \end{tabular}
  \end{center}
\end{table}

% some present result
We have carefully evaluated instrumental energy scales of the \suzaku XIS, 
used iron K-shell line emission, 
and found a radial velocity shift of the second gas component with respect to the main cluster.
The X-ray determined redshifts are consistent with those of galaxy components.
This is the most robust detection of a gas bulk motion in a cluster.

Throughout this paper,
we assume cosmological parameters as follows; $H_0 = 70$ km s$^{-1}$Mpc$^{-1}$, 
$\Omega_\mathrm{m} = 0.3$, and $\Omega_\mathrm{\Lambda} = 0.7$.
% http://heasarc.nasa.gov/xanadu/xspec/manual/XScosmo.html
At the cluster redshift of 0.058, 
% from Sun 2002
one arc-minute corresponds to 67.4~kpc.
% http://www.astro.ucla.edu/~wright/CosmoCalc.html
We use the 68\% ($1\sigma$) confidence level for errors, unless stated otherwise.

\section{Observations}\label{sect:obs}
\begin{figure}
\begin{center}
    \FigureFile(80mm,80mm){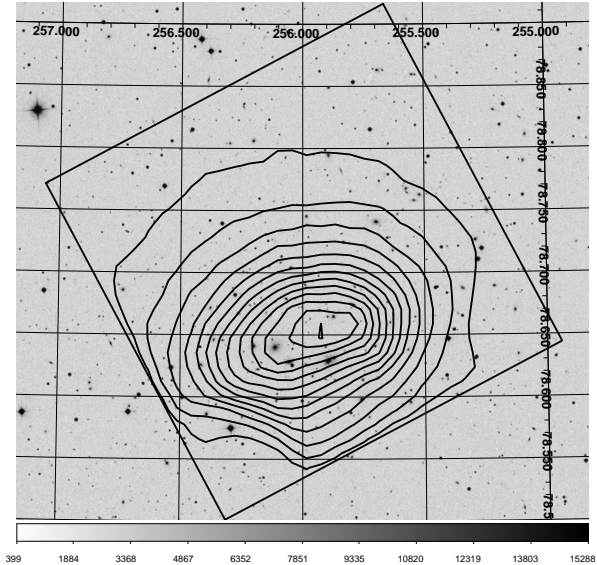}
\end{center}
\caption{\suzaku XIS contour image of A~2256 in the 0.3--10 keV band
overlaid on a DSS (Digital Sky Survey by the Space Telescope Science Institute) optical image.
The three XIS data are co-added.
The XIS field of view is shown by a square.
The contour levels are linear from 10 to 160 counts pixel$^{-1}$.
No vignetting nor background was corrected.
The calibration-source regions at corners of CCD are excluded.
North is up and east is to left.
}
\label{fig:sky-image}
\end{figure}

\suzaku observations of A~2256 were performed on 2006 November 10--13 (PI: K. Hayashida).
The XIS was in the normal window and the spaced-row charge injection off modes.
% The effective exposure time was 94.4~ks.
The observation log is shown in table~\ref{tbl:obs}.
Figure ~\ref{fig:sky-image} shows an X-ray image of the cluster.
Detailed descriptions of the \suzaku observatory, the XIS instrument, 
and the X-ray telescope are found in 
\citet{mitsuda07}, \citet{koyama07}, \citet{serlemitsos07}, respectively.

To verify the XIS gain calibration, 
we have also used data from the Perseus cluster observed in 2006 with the same XIS modes as those for A~2256 (table~\ref{tbl:obs}).
These data were already used for the XIS calibration (e.g. \cite{ozawa09}) and scientific analyses (\cite{tamura09}; \cite{nishino10}).

\begin{table*}[h]
  \caption{{\it Suzaku} observations of A~2256 and the Perseus cluster. }
\label{tbl:obs}
  \begin{center}
    \begin{tabular}{llllll}
\hline \hline
Target & Date & Sequence    & (RA, Dec) & P.A. \footnotemark[$*$] & Net exposure\\
     &      & number   & (degree, J2000.0) & (degree) & (ks) \\
\hline
A~2256 & 2006/11/10--13 & 801061010 & (256.0138, 78.7112) & 208 & 94.4 \\
Perseus & 2006/2/1 & 800010010 & (49.9436, 41.5175) & 260 & 43.7 \\
Perseus & 2006/8/29 & 101012010 & (49.9554, 41.5039) & 66 & 46.6  \\
% RA, Dec from Suzaku master
\hline
\multicolumn{6}{@{}l@{}}{\hbox to 0pt{\parbox{85mm}{\footnotesize
\par\noindent
\footnotemark[$*$] The nominal position angle (north to DETY axis).
}\hss}}
    \end{tabular}
  \end{center}

\end{table*}

%% 2010-8-6

\section{Analysis and Results}
\subsection{Data Reduction}
We used version 2.1 processing data along with the HEASOFT version 6.9.
% i02111@klm~/KLM/suzaku/a2256/20100708/reg-0804$ls -l /darts/suzaku/ver2/801061010
% lrwxrwxrwx  1 21009 21009 19 Jul 20 10:49 /darts/suzaku/ver2/801061010 -> ../ver2.1/801061010
In addition to the XIS standard event selection criteria, 
we have screened data by adding the following conditions;
geomagnetic cut-off-rigidity $>$ 6~GV, 
the elevation angle above the sunlit earth $>$\timeform{20D} and the dark earth $>$\timeform{5D}.
%~/backup/klm/suzaku/a2256/20100708/reg-0804/xis0-mainR240.log
We used the latest calibration file as of July 2010.
Using these files, we corrected the XIS energy scale.
% i02111@klm~/KLM/suzaku/a2256/20100708/reg-0804$ls -l -t /usr/local/xray/caldb/data/suzaku/xis/index/  | head
% -rw-r--r--  1 20002 20002 1077120 Jul  9 19:42 caldb.indx20100709
The data from three CCDs (XIS 0, XIS 1, and XIS 3) are used.

We examined the light curve excluding the central bright region events ($R<6'$) for stable-background periods.
There was no flaring event in the data.
% (nxb)
The instrumental (non-X-ray) background was estimated using the night earth observation database and the software {\tt xisnxbgen} \citep{tawa08}.

% (arf, rmf)
We prepared X-ray telescope and CCD response functions for each spectrum using software {\tt xissimarfgen} \citep{ishisaki07} 
and {\tt xisrmfgen} \citep{ishisaki07}, respectively.
The energy bin size is 2~eV.
% note-p.189

To describe the thermal emission from a collisional ionization equilibrium (CIE) plasma, 
we use the APEC model \citep{smith01} with the solar metal abundances
taken from \citet{ag89}.
% http://heasarc.gsfc.nasa.gov/docs/xanadu/xspec/manual/XSabund.html
% O/H = 8.51e-04, Fe/H=3.63e-04
% The Galactic absorption is described by the photo-electric absorption of Wisconsin cross-sections (wabs model in XSPEC)
% with the reported neutral hydrogen column density of $1.5\times 10^{-21}$cm$^{-2}$ toward the cluster.
%http://heasarc.nasa.gov/xanadu/xspec/manual/XSmodelWabs.html

\subsection{Energy Scale Calibration}\label{sect:cal}
The main purpose of this paper is to measure Doppler shifts of K-shell iron lines in X-ray.
In these analyses, the energy-scale calibration is crucial.
In \S~\ref{sect:cal-status}, 
we summarise the calibration status.
In \S~\ref{sect:cal-cal},
we attempt to confirm the calibration using calibration-source data collected during the A~2256 observation.
The primal goal here is to find differences of line energies observed in different positions within the field of view. 
Therefore the positional dependence of the energy scale is most important, 
which is evaluated in \S~\ref{sect:cal-per}.
Here we focus on the data obtained in spaced-row charge injection off mode
and around the K-shell iron lines.
Considering all available information given here as summarized in table~\ref{tbl:gain-cal}, 
we assume that the systematic uncertainty of the energy scale around the iron lines is most likely 0.1\% 
and 0.2\% at most over the central $14'.7 \times 14'.7$ region or among the three CCDs.

\subsubsection{Reported Status}\label{sect:cal-status}
\citet{koyama07b} estimated the systematic uncertainty of the absolute energy in the iron band to be within $+0.1,-0.05$\%, 
based on the observed lines from the Galactic center along with 
Mn K$\alpha$ and K$\beta$ lines (at 5895~eV and 6490~eV respectively) from the built-in calibration source ($^{55}$Fe).
Independently, 
\citet{ota07} investigated the XIS data of two bright and extended sources (the A~1060 and Perseus clusters) 
and evaluated the positional energy scale calibration in detail.
They estimated the systematic error of the spatial gain non-uniformity to be $\pm 0.13$\%.
% A1060, 2005-11, Per 2006-Feb
Furthermore, 
\citet{ozawa09} examined systematically the XIS data obtained from the start of operation in July 2005 until December 2006.
% and used checker flag charge-injection technique.
They reported that position dependence of the energy scale are well corrected for the charge-transfer inefficiency and that the time-averaged uncertainty of the absolute energy is $\pm$ 0.1\%.
In addition, the gradual change of the energy resolution is also calibrated;
the typical uncertainty of resolutions is 10--20~eV in full width half maximum.

\begin{table*}[h]
  \caption{Accuracy of the XIS energy scale calibration and the observed velocity shift}
\label{tbl:gain-cal}
  \begin{center}
    \begin{tabular}{llll}
\hline \hline
Section$\dagger$ & Line used & $\Delta *$ (\%) & Reference and remarks \\
\hline
\ref{sect:cal-status}& 
GCDX/Fe-K$\ddagger$ & $+0.1, -0.05$ & \citet{koyama07b}; uncertainty of the absolute energy. \\
\ref{sect:cal-status}& 
Perseus/Fe-K     & $\pm0.13$ & \citet{ota07}; spatial non-uniformity (at 68\% confidence). \\
\ref{sect:cal-status}& 
Mn I K${\alpha}$ & $\pm0.1$ & \citet{ozawa09}; time-averaged uncertainty of the absolute energy. \\
\ref{sect:cal-cal}& 
Mn I K${\alpha}$ & $\pm0.1$ & This paper; standard deviation among CCD segments in the A~2256 data. \\
\ref{sect:cal-per}& 
Perseus/Fe-K & (-0.09--0.06) & This paper; variations among the two regions.\\
\hline
\ref{sect:vel}& 
A~2256/Fe-K & $0.5\pm 0.1$ & This paper; velocity shift in A~2256, after correcting the inter-CCD gain \\
&           &               & and the statistical error. \\
\hline
\multicolumn{4}{l}{ \parbox{150mm}{\footnotesize
\footnotemark[$\dagger$] Section in this paper.\\
\footnotemark[$*$] 
possible uncertainty, velocity shift, or its error all in percentage of the line energy. 
At the iron K line, 0.1\% corresponds to 7~eV for the energy shift 
or 300 km s$^{-1}$ for the radial velocity.\\
\footnotemark[$\ddagger$] From the Galactic center diffuse emission.\\
}}
    \end{tabular}
  \end{center}
\end{table*}

\subsubsection{Absolute Scale}\label{sect:cal-cal}
% 2010-8 p.161, 2010-12-1, p.29
We extracted spectra of calibration sources which illuminate two corners of each CCD (segment A and D). 
These spectra in the energy range of 5.3--7.0~keV 
are fitted with two Gaussian lines for the Mn K${\alpha}$ and K${\beta}$ along with a bremsstrahlung continuum component.
Here we fixed the energy ratio between the two lines to the expected one.
Thus obtained energy centroids of the Mn K${\alpha}$ line from  the two corners of three CCDs
give an average of 5904~eV (as compared with the expected value of 5895~eV) and a standard deviation (scatter among the six centroids) of 6~eV.
The statistical errors of the line center were about 1--2~eV.
This confirms that the absolute energy scale averaged over all CCDs
and the relative gain among CCD segments are within $\pm 0.15$\% and $\pm 0.10$\%, respectively.
%% (e.g. 5904/5895 $\sim$ 1.0015).
Simultaneously we found that the data can be fitted with no intrinsic line width for the Gaussian components, 
meaning that the energy resolution is also well calibrated. 
% note 2010-8-4, cal-e.pl
% XIS0-A,5906~eV ;
% XIS0-D,5912~eV ;
% XIS1-A,5895~eV ;
% XIS1-D,5902~eV ;
% XIS0-A,5910~eV ;
% XIS0-D,5906~eV.

\subsubsection{Spatial Variation}\label{sect:cal-per}
%% 2010-12-10
We used the XIS data of the Perseus cluster, 
which would provide the highest-quality XIS line spectra over the whole CCD field of view
among all the \suzaku observations.
Here we assume no line shift intrinsic to the cluster within the observed region.
There were two exposures of the Perseus in the normal window and the spaced-row charge injection off modes (table ~\ref{tbl:obs}) in periods close to the A~2256 observation.
Note that \citet{ota07} used the same data, but with early calibration (i.e. version 0.7 data).
Here we re-examine the accuracy with latest and improved calibration.

Firstly, following \citet{ota07}, we divided the XIS field of view into $8 \times 8$ cells of size $2'.1\times2'.1$.
Each spectra in the 6.2--7.0~keV band is fitted with two Gaussian lines 
for He-like K${\alpha}$ ($\sim 6700$~eV) and H-like K${\alpha}$ ($\sim 6966$~eV)
and a bremsstrahlung continuum model.
Here we fixed the ratio of the central energies between the two lines to a model value of 1.040
and let the energy of the first line be a fitting parameter.
Because of the low statistics data in CCD peripheral regions
we focus on the central $7 \times 7$ cells (i.e., $14'.7 \times 14'.7$).
The typical statistical errors of the line energy are from $\pm 4$~eV to $\pm 25$~eV.
Thus derived central energies from $7 \times 7$ cells  for each CCD
are used to derived an average and a standard deviation.
The average values are 6575~eV, 6575~eV, and 6569~eV, for XIS 0, XIS 1 and XIS 3, respectively,
which are consistent with the cluster redshift of 0.017--0.018.
The standard deviations among $7 \times 7$ cells are 7~eV, 13~eV, and 10~eV  for the three XISs, respectively.
There is no cell having a significant deviation from the average value at more than 2$\sigma$ level.
% to-do see note p.43 for more accurate value, 2011-1-13
These deviations include not only systematics due to the instrumental gain uncertainty
but also statistics and systematics intrinsic to the Perseus emission.
Therefore 
the energy scale uncertainty should be smaller than this range of deviations (0.1--0.2\%).

% /home/cc/i02111/work/suzaku/perseus/201012cal
%%% 2010-12-1
Secondly, we focus on the CCD regions specific in our analysis below.
We have extracted the Perseus spectra from the same detector regions (main and sub) as used in the A~2256 analysis.
The definition of the regions are given in figure~\ref{fig:fe-image} and \S~\ref{sect:vel}.
% (two exposures $\times$ two regions $\times$ three CCDs)
These spectra were fitted with the same model as above and used to derive the redshift (from the line centroid).
We found no difference in redshift between the two regions.
The redshift differences (between the two regions from the three sensors) range 
from $-9\times 10^{-4}$ to $+6\times 10^{-4}$ 
with an average of $1.5\times 10^{-5}$.
Accordingly we estimate the possible instrumental gain shift to be within $\pm$0.1\% 
with no systematic bias between the two regions.

\subsection{Energy sorted X-ray Images}\label{sect:e-image}
\begin{figure}
\begin{center}
    \FigureFile(180mm,80mm){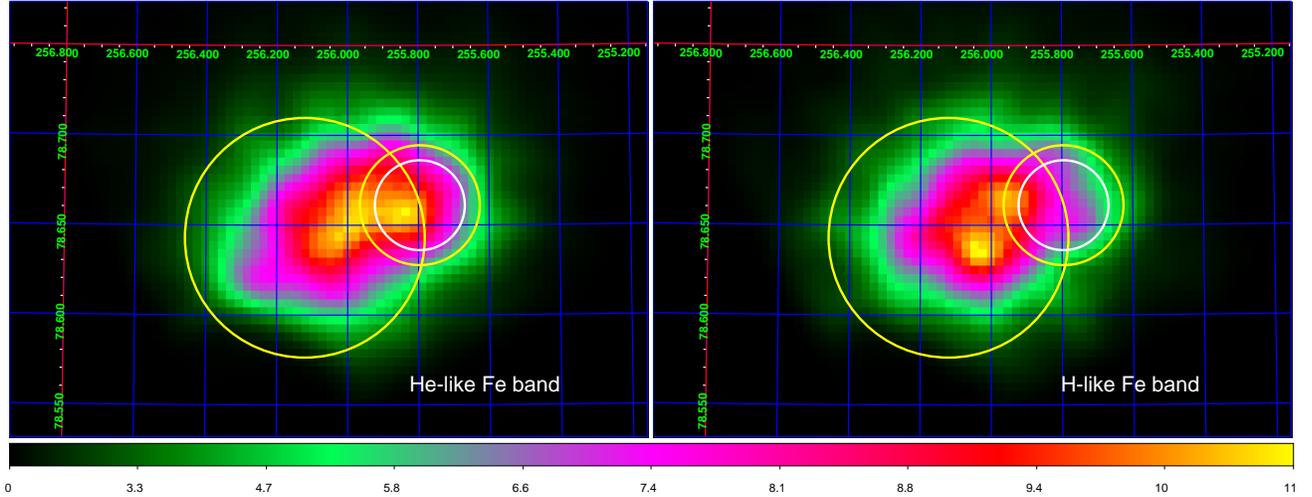}
\end{center}
\caption{XIS images of A~2256 
from the 6.10--6.48~keV energy band including He-like iron emission (left)
and from the 6.48--6.80~keV band including H-like iron emission (right) in unit of counts.
No vignetting nor background was corrected.
Images have been smoothed with a Gaussian filter with $\sigma=17''$.
The spectra are extracted from the two regions indicated by circles in yellow (the main region)
and a circle in white (the sub).
}
\label{fig:fe-image}
\end{figure}

In order to examine the spectral variation over the A~2256 central region, 
we extracted two images in different energy bands including He-like and H-like iron line emission 
as shown in figure~\ref{fig:fe-image}.
There appear at least two emission components which corresponds to the main cluster at east and the sub component at west
discovered in \citet{briel91}.
We noticed a clear difference in the distribution between the two images.
In the He-like iron image, the sub component exhibits brightness comparable to that of the main component.
On the other hand, in the H-like iron image, the sub has lower brightness.
This clearly indicates that the sub has cooler emission compared with the main.
Based on these contrasting spatial distributions, we define centers of the two components
to be (\timeform{256D.1208}, \timeform{78D.6431}) and (\timeform{255D.7958}, \timeform{78D.6611}), 
in equatorial J2000.0 coordinates (RA, Dec), respectively, as shown in figure~\ref{fig:fe-image}.
% see positions.txt for detail
%These positions correspond the P$_{1}$ and P$_{2}$ in \citet{sun02}.
%In these images, we indicate regions of the main and sub components, as defined and used below.

%% 2010-12-13, see Notep.48

\subsection{Radial Velocity Shift}\label{sect:vel}
\subsubsection{Separate Spectral Fitting}
Our goal is to constrain the velocity shift of the sub component with respect to the main cluster.
Then, we extracted sets of spectra from the two components  and fitted with different redshifts.
More specifically,
for the main component emission, we integrated the data within $4'$ in radius from the main center 
but excluding a sub component region with a radius of $2'$ (figure\ref{fig:fe-image}). 
For the sub region, data within $1'.5$ in radius from the sub center are extracted.
% See figure~\ref{fig:fe-image}) for these main and sub regions.

We use the energy range of the 5.5--7.3~keV around the iron line complex.
In this energy band the cosmic background fluxes are below a few \% of the source count in the two regions.
Therefore we ignore this background contribution.

\begin{figure}
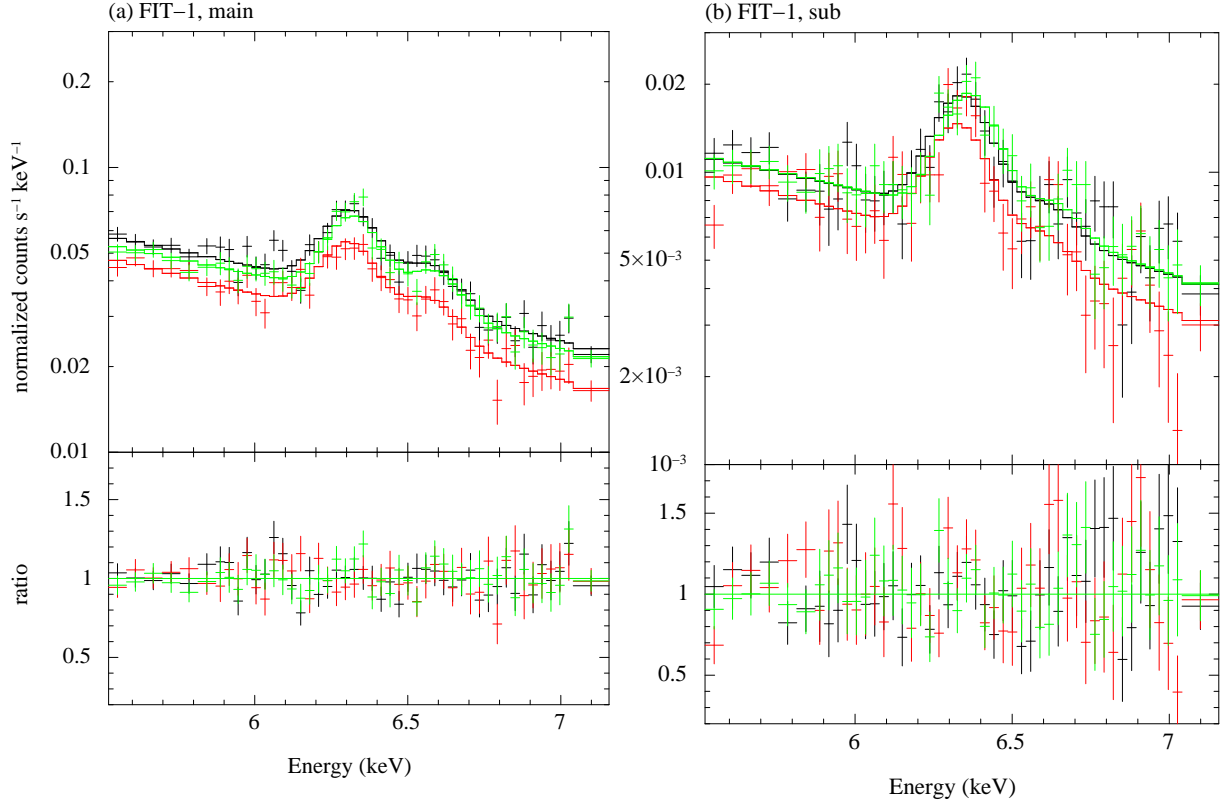

\begin{center}
    \FigureFile(80mm,80mm){figure3a.ps}
    \FigureFile(80mm,80mm){figure3b.ps}
\end{center}
% (reg-0804v2/fit2) 
\caption{
Cluster spectra along with the best-fit CIE model in histogram (FIT-1).
Plots (a) and (b) are from the main and sub regions, respectively.
The XIS 0, XIS 1, and XIS 3 data are shown by black, red, and green colors, respectively.
In lower panels fit residuals in terms of the data to model ratio are shown.
}
\label{fig:fit2-sp}
\end{figure}

%%% 2010-11-30, reg-0804v2/fit2
%%% (** reg-0804v2/fit2)\\
We use a CIE component to model the spectra.
Free parameters are an iron line redshift,
a temperature, an iron abundance, and normalization. 
Models for the three XISs are assumed to have different redshifts and normalization 
to compensate inter-CCD calibration uncertainties.
Other parameters, temperature and abundance, are fixed to be common among the CCDs.
As shown in figure~\ref{fig:fit2-sp} and table~\ref{tbl:fit1}, 
these models describe the data well (FIT-1).
The fits give different redshifts between the two regions from all three XISs as shown in figure~\ref{fig:z1}(a).
The statistical errors of the absolute redshift are less than $\pm 0.002$.
The redshift differences are 0.0048, 0.0041, and 0.0076 for the three XIS.
These are equivalent to shifts of 30--50~eV in energy or 1200--2300 km s$^{-1}$ in radial velocity.

%%% reg-0804v2/fitg1
%% (** reg-0804v2/fitg1)\\
The spectral shape of the two regions are different as shown in figure~\ref{fig:fit2-sp}.
Therefore, the obtained redshift difference may depend on the spectral modeling.
To check this possibility, 
instead of the CIE model, we use two Gaussian lines and a bremsstrahlung continuum component (FIT-2).
Here we assume the He-like iron resonance line centered on 6682~eV
and the H-like iron Ly$\alpha$ centered on 6965~eV
for the two lines 
and determine the redshift common to these two lines.
The Gaussian components are set to have no intrinsic width.
As given in table~\ref{tbl:fit1}, these models give slightly better fits to the data in general
than the first model .
For each XIS, we obtained redshifts and therefore a redshift difference 
as same as those from the first fit within the statistical uncertainty.
Therefore the redshift depends insignificantly on the spectral modeling.

% 2010-8-11, note-p184
Strictly speaking line centroids of the iron line transitions 
could change depending on the emission temperature.
In the case of the observed region of A~2256, 
where the temperature varies within 4--8~keV (e.g. \cite{sun02}),
the strong iron line structure is dominated by the He-like triplet.
Within this temperature range, 
the emission centroid of the triplet stays within 6682--6684~eV based on the APEC model.
This possible shift ($< 2$~eV) is well below the obtained redshift difference (30--50~eV)
and should not be the main origin of the difference.

\subsubsection{Gain-corrected Spectral Fitting}
\begin{figure}
\begin{center}
% see cpImage.sh, reg-0804v2/
%~/A2256/backup/20100708/reg-0804v2/z1.qdp
  \FigureFile(80mm,80mm){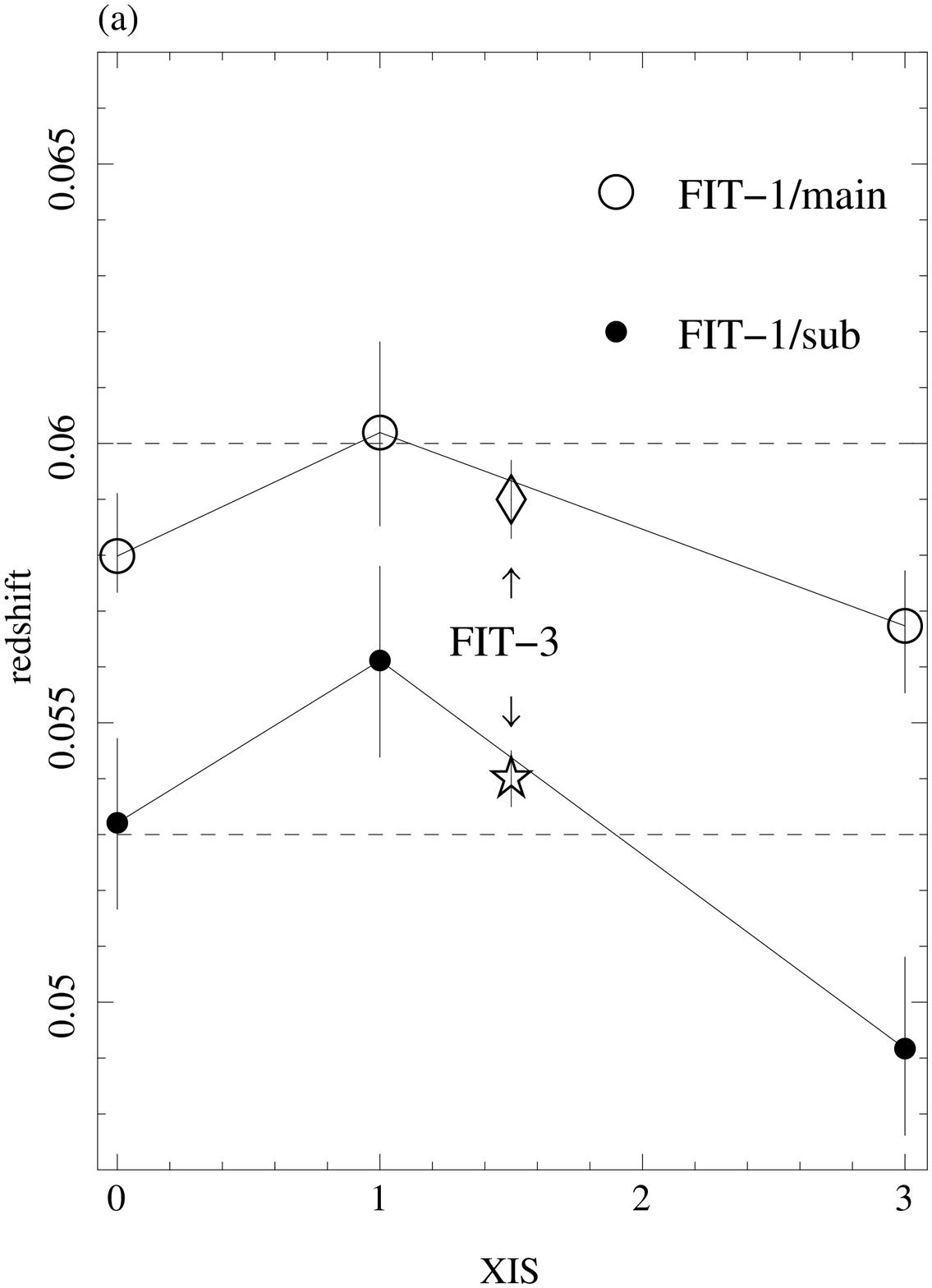}
  \FigureFile(80mm,80mm){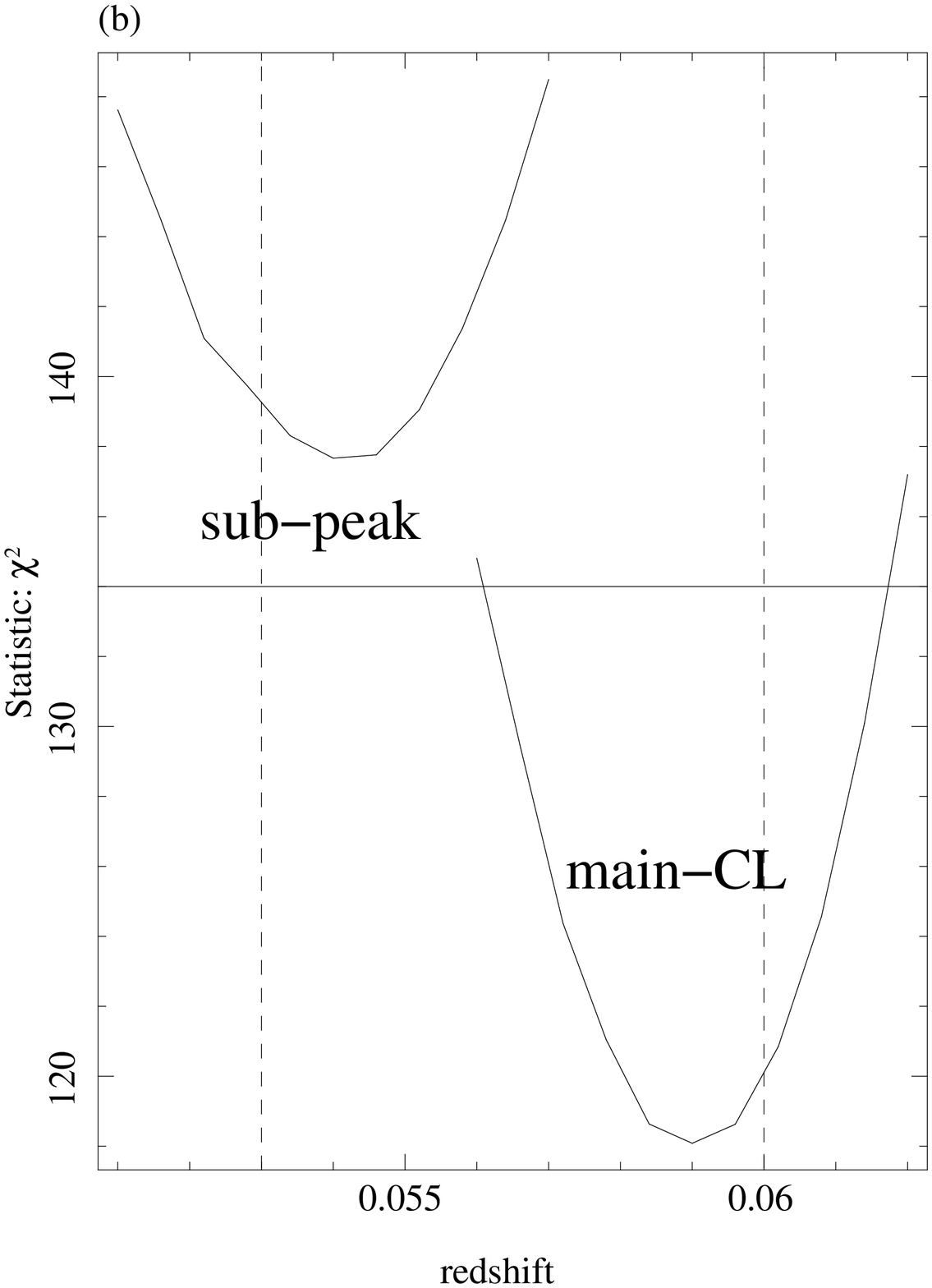}
%  \FigureFile(70mm,70mm){figure4a.ps}
%  \FigureFile(70mm,70mm){figure4b.ps}
\end{center}
\caption{(a) The redshifts obtained from the CIE fitting (FIT-1) of the main and sub component regions for each XIS (X-axis). 
The gain-corrected results of FIT-3 are shown at the position of XIS$=$1.5.
The two horizontal dashed lines indicate redshifts from the optical data (\cite{briel91}; table~\ref{tbl:velocity}).
(b) The statistic $\chi^2$ distributions of the gain-corrected Gaussian fitting (FIT-3)
of the two regions as a function of the redshift.
The horizontal line at $\chi^2=134$ indicates the degree of the freedom of the fits.
The two vertical dashed lines indicate the redshifts as in (a).
}
\label{fig:z1}
\end{figure}

% 2010-8-10
% (** reg-0804v2/fitg2)\\
The obtained redshifts are systematically different among the three XISs [figure~\ref{fig:z1}(a)].
We attempt to correct this inter-CCD gain difference based on the calibration source data.
Following \citet{fujita08}, 
we estimate a gain correction factor, $f_{\rm gain}$, by dividing the obtained energy of the Mn K line from the calibration source by the expected one.
In the A~2256 data as given in section \ref{sect:cal}, 
$f_{\rm gain}$ are 1.0018 (XIS 0), 1.0000 (XIS 1), and 1.0016 (XIS 3).
This factor, the redshift obtained from the fit ($z_{\rm fit}$), and the corrected redshift ($z_{\rm cor}$)
have a relation 
\begin{equation}
f_{\rm gain} = \frac{z_{\rm cor}+1}{z_{\rm fit}+1}.
\end{equation}
We use this correction and fit the spectra with a single redshift common to the three XISs
to the two-Gaussian model (FIT-3).
These models give slightly poorer fits compared with the previous ones,
because of a decrease of the degree of freedom
(table~\ref{tbl:fit1}). 
Nevertheless, the fits are still acceptable.
The difference in the redshift between the two region
is $0.005\pm 0.0008$ (or $1500\pm240$ km s$^{-1}$).
%% see note.171, 2010-8
In figure~\ref{fig:z1}(b) we show the statistical $\chi^2$ distribution as a function of the redshift.
This indicates clearly that the data cannot be described by the same redshift for the main and sub regions.

% 2011-1-31
We found that redshifts determined here by X-ray are consistent with those of member galaxies in optical (table~\ref{tbl:velocity}) as explained below.
The X-ray redshift of the main component, z=0.0059 or 17700 km s$^{-1}$, 
is the same as the galaxy redshift within the statistical error.
That of the sub component, z=0.00540 or 16200 km s$^{-1}$, 
is larger than the galaxy value of 15730 km s$^{-1}$.
Yet the difference, 470 km s$^{-1}$, is within the combined errors from X-ray statistical (150 km s$^{-1}$), 
systematic (300km s$^{-1}$), and optical fitting (160 km s$^{-1}$).
Besides this, 
the obtained specta of the sub component could be contaminated from the emission of the main component due to the projection and the telescope point spread function (half power diameter of about $2'$).
This contamination could make the obtained redshift of the sub larger than the true one (or closer to that of the main).
 
\begin{figure}
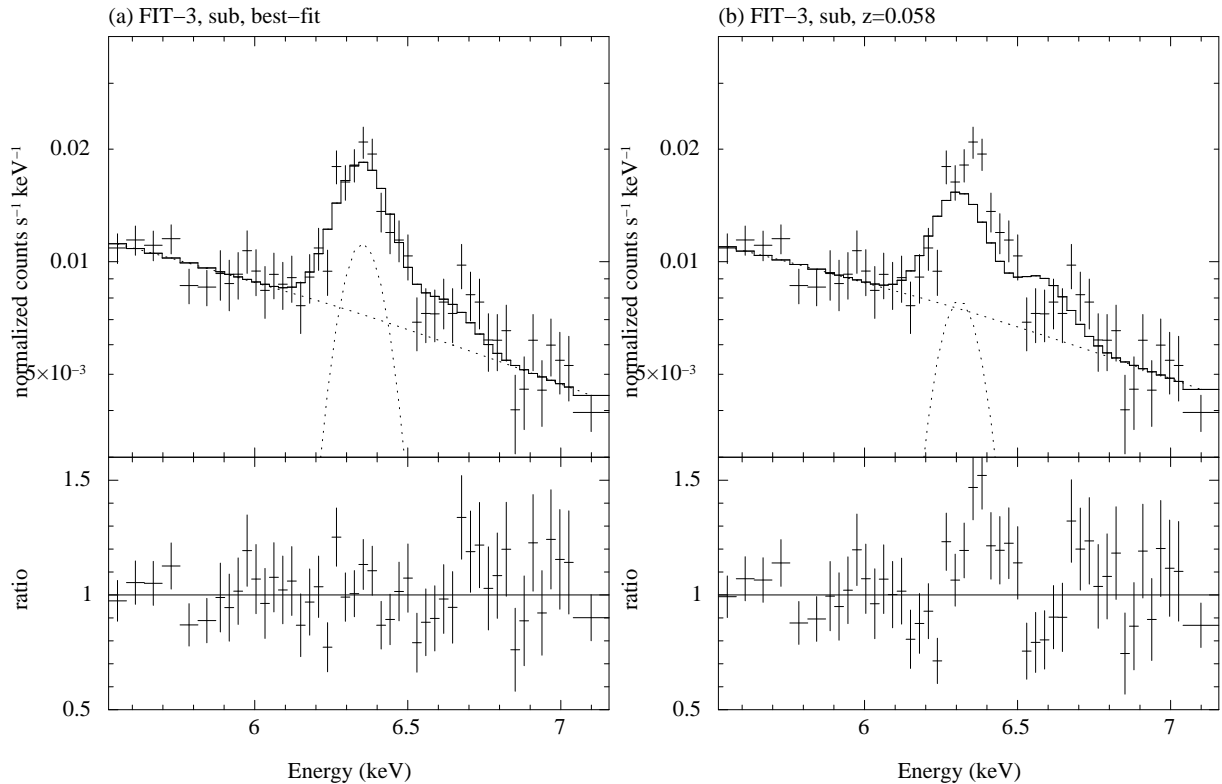

\begin{center}
  \FigureFile(80mm,80mm){figure5a.ps}
  \FigureFile(80mm,80mm){figure5b.ps}
\end{center}
\caption{The spectra from the sub-component region by adding the two front-illuminated XIS (0 and 3),
fitted with two Gaussians and a continuum model.
(a) The redshift common to the two lines is left free.
The model gives the best-fit redshift of 0.05115 and $\chi^2$ of 40.4 for the degrees of the freedom of 42.
(b) The redshift is fixed to the main component value ($z=0.058$).
The model gives $\chi^2$ of 80.8 for degrees of freedom of 43.}
\label{fig:1d-fit}
\end{figure}

% 2010-8-11
To visualize the redshift shift in the sub region more directly, 
we show spectral fittings by combining the two front-illuminated XIS (0 and 3) data
in figure~\ref{fig:1d-fit}.
We fitted the spectra with the two Gaussian model as given above.
In this figure we compare the fitting results between the best-fit model ($z=0.051$)
and one with the redshift fixed to the main cluster value ($z=0.058$).
This comparison shows that we can reject that the sub region has the same redshift as the main cluster.

%% 2010-8-6
% i02111@klm~/work/suzaku/a2256/20100708/reg-0804v2$fit2-main1.sh
% i02111@klm~/work/suzaku/a2256/20100708/reg-0804v2$fit2-sub1.sh

%% 2010-8-6

%% 2010-8-11

\begin{table}[p]
%\begin{table}
\caption{Spectral fitting results from the main and sub regions.}
\label{tbl:fit1}
  \begin{center}
    \begin{tabular}{lllll}
\hline \hline
Region  & $kT$ & Fe & $<z>$ $*$ & $\chi^2/d.o.f.$ \\
       & (keV) & (Solar)  & & \\
\hline  
% fit2, mainR240_log2.log, mainR240_log2.log, 2010-11-30
% ~/backup/klm/work/i02111/suzaku/a2256/20100708/reg-0804v2/fit2/
\multicolumn{5}{c}{FIT-1, CIE component}\\
main & 7.1$\pm0.3$ & 0.25$\pm0.02$ & 0.0583 & 121/139 \\
sub  & 5.0$\pm0.4$ & 0.29$\pm0.03$ & 0.0528 & 135/139 \\
\hline
\multicolumn{5}{c}{FIT-2, two Gaussian components}\\
% ~/backup/klm/work/i02111/suzaku/a2256/20100708/reg-0804v2/fitg1/
main & 6.2$\pm0.7$ & - & 0.0580 & 112/137 \\
sub  & 4.5$\pm0.9$ & - & 0.0528 & 131/137 \\
\hline
\multicolumn{5}{c}{FIT-3, two Gaussian with the gain correction}\\
% ~/backup/klm/work/i02111/suzaku/a2256/20100708/reg-0804v2/fitg2/
main & 6.2$\pm0.7$ & - & 0.0590$\pm0.0007$ & 118/137 \\
sub  & 4.6$\pm0.9$ & - & 0.0540$\pm0.0005$ & 137/137 \\
\hline
\multicolumn{5}{@{}l@{}}{\hbox to 0pt{\parbox{85mm}{\footnotesize
\par\noindent
\footnotemark[$*$] Average value over three XIS data for the FIT-1 and FIT-2 are given.
Statistical errors on these values are typically 0.001.
In the FIT-3, a single redshift common to the three XISs is assumed.
\\
%%% 2010-12-13 note.47
}\hss}}
    \end{tabular}
  \end{center}
\end{table}

\section{Summary and Discussion}
\subsection{Summary of the Result}
% \today \\
%(What is new result based on the data)\\
% 結果で示したデータに元に、こういうことを明らかにしたという主張\\
We found gas bulk motion of the second component in A~2256.
The difference in the redshifts and hence radial velocities between the main and sub systems
is found to be 1500 $\pm 300$ (statistical)  $\pm 300$ (systematic) km s$^{-1}$ (\S~\ref{sect:vel}).
This observed shift corresponds to only 0.5\% in the energy,
but is well beyond the accuracy of the energy scale reported by the instrument team (\cite{koyama07b}; \cite{ozawa09})
and that by \citet{ota07}.
Focusing on the present analysis of A~2256, 
we also have examined the calibration systematics independently 
and confirmed the accuracy given above.
The obtained redshifts and hence the difference between the two X-ray emitting gas components
are consistent with those of the radial velocity distribution in member galaxies.
This consistency uniquely strengthens the reliability of our X-ray measurement.

This is the fist detection of gas bulk motion
not only in A~2256 but also from \suzaku,
which presumably has the best X-ray spectrometer in operation for the velocity measurement.
As given in \S~\ref{sect:intro}, Dupke and Bregman (2001a, 2001b, 2006) previously claimed detections of bulk motions in the Perseus and Centaurus clusters. 
Compared with these and other attempts,
our measurement is more accurate and robust.
%% ??? compare errors ???
This improvement is not only due to the better sensitivity and calibration of the XIS
but also due to the well-separated and X-ray bright nature of the structure in A~2256.

% (how contribute to the problems given in intro,
% improvement over the previos result)\\
% 2. introで提起した問題の解決に向けてどう貢献したか
% 今までの研究
%- それに比べて、この研究はどう進歩しているか
% comparison with Dupke results.\\

Radial velocity distributions of cluster galaxies
were used to evaluate dynamics of cluster mergers.
These studies, however, have limited capability.
First, largely because of the finite number of galaxies and the projection effect, 
a galaxy sub group within a main cluster is not straightforward to identify.
Second, the major baryon in the cluster is not galaxies but the gas in most systems.
Therefore dynamical energy in the hot gas cannot be ignored.
Third, galaxies and gas are not necessary to move together especially during merging phases,
because of the different nature of the two (collisionless galaxies and collisional gas).
In practice, some clusters under the violent merging phase such as 1E~0657-558 (\cite{clowe06}) 
and A~754 (\cite{mar03}) show a spatial separation between the galaxy and gas distributions.
Therefore, it is important to measure dynamics of galaxies and gas simultaneously.
The present result is one of the first such attempts.

\subsection{Dynamics of A~2256}
%%% 2011-1-25
The determined radial velocity, $v_r\sim 1500$~km s$^{-1}$, 
gives an lower limit of a three dimensional true velocity, $v = v_r/\sin\alpha$, 
where $\alpha$ denotes the angle between the motion and the plane of the sky.
Given a gas temperature of 5~keV (\S~\ref{sect:vel}) and an equivalent sound speed of 1100~km s$^{-1}$ around the sub component, this velocity corresponds to a Mach number $M> 1.4$.
Therefore, at least around the sub component, 
the gas kinematic pressure (or energy) can be $(1.4)^2\sim 2 $ times larger than the thermal one. 
In this environment, the gas departs from hydrostatic equilibrium.
Then, does this motion affect the estimation of the mass of the primary cluster ?
As argued by \citet{mar97}, it depends on the physical separation between the two components.
In the case of A~2256, the two are not likely too closely connected 
to disturb the hydrostatic condition around the primary,
as estimated below.
However, to weight the total mass within the larger volume including the sub component 
we should consider not only the mass of the sub itself but also the dynamical pressure associated with the relative motion.
This kind of a departure from hydrostatic equilibrium was predicted 
generally at cluster outer regions in numerical simulations (e.g. \cite{evrard96}).

We will compare our measurement with other studies on the ICM dynamics.
\citet{m02} discovered a bow shock in the \chandra image of 1E~0657-56 (the bullet cluster) and
estimated its $M$ to be $3.0\pm 0.4$ (4700 km~s$^{-1}$) 
based on the observed density jump under the shock front condition.
Using a similar analysis, \citet{m05} derived $M$ of $2.1\pm0.4$ (2300 km~s$^{-1}$) in A~520.
Based on a more specific configuration in the ``cold front'' in A~3367, 
\citet{v01} estimated its $M$ to be $1.0\pm 0.2$ (1400 km~s$^{-1}$).
In these cases, velocity directions are assumed to be in the plane of the sky.
See \citet{mv07} for a detailed discussion and other examples.
These measurements are unique but require certain configurations of the collision to apply to observations.
In contrast, 
X-ray Doppler shift measurements as given in the present paper
are more direct and commonly applicable to merging systems.
This method is sensitive to the motion parallel to the line of sight.
These two measurements are complementary and could provide a direct measurement of three dimensional motion
if applied to a merging system simultaneously.

%(implication for more general issues)\\
% その問題を解決したことで得られる、より一般的な学術的な意義
% 2011-1-11
We measured the gas velocity parallel to the line of sight ($v_r$).
How about the velocity in the plane of the sky ($v_{\rm sky}$) and the true velocity $v$?
The \rosat X-ray image of A~2256 shows a steeper brightness gradient between the main and second components \citep{briel91}.
Furthermore, from the \chandra spectroscopic data \citet{sun02} found a hint of a temperature jump across the two component.
They argued a similarity of this feature to the ``cold front'' discovered in other clusters.
The temperature and hence pressure jump in A~2256 (from about 4.5~keV to 8.5~keV)
are similar to those found in A~3367, 
in which the jump indicates a gas motion with $v_{\rm sky}$ of about 1400~km~s$^{-1}$.
Therefore we expect a similar $v_{\rm sky}$ in A~2256, 
which is comparable to $v_r$.
Further assuming that the second component directs to the main cluster center, 
$v$ is estimated as 
$\sqrt{v_r^2+v_{\rm sky}^2} \sim \sqrt{2}v_r \sim 2000$~km s$^{-1}$.
Instead of assuming $v_{\rm sky}$, by considering the mean $\sin\alpha$ factor, $2/\pi$, 
$v$ becomes 2400~km s$^{-1}$ on average.
Based on a simple assumption that the two system started to collide from rest, 
we can estimate the velocity as 
$v \sim (\frac{2GM}{R})^{1/2}$, 
where $G, M$ and $R$ are the gravitational constant, the total mass of the system, and the separation, respectively.
The total mass of A~2256 of $8\times 10^{14}$~M$_{\odot}$ (\cite{mar97})
and an assumed final separation $R$ of 1~Mpc
give $v \sim 2800$~km s$^{-1}$, which is comparable but larger than the estimated current velocity.
Putting $v = 2000-2400$~km s$^{-1}$ to the relation, 
we obtain a current separation $R$ of $1.4-2$~Mpc.
The time to the final collision can be estimated to be about 0.2--0.4~Gyr [$(R-1)$~Mpc divided by $v$].
Here we assume that the system is going to the final collision.
This assumption is consistent with a lack of evidence for strong disturbances in the X-ray structure, 
as argued previously for A~2256 in general.

\subsection{Future Prospect}
%(what should be do in the next step)\\
% 今後の発展
% - 問題を完全解決するためには、今後何を行うべきか
% - 次に取り組みべき問題は
Our observations provided the first detection of gas bulk motion in a cluster.
To understand in general the cluster formation which is dominated by non-linear processes, 
systematic measurements in a sample of clusters are required.
For example, some clusters such as 1E~0657-558 (\cite{clowe06}) at violent merging stages 
show segregation of the gas from the galaxy and possibility from dark matter components.
In these systems, we expect different situations in the gas and galaxy dynamics compared with that found in A~2256.
Given capabilities of current X-ray instruments such as the \suzaku XIS, 
A~2256 is presumably an unique target with X-ray flux high and the velocity separation clear enough to resolve the structure.
Accordingly, the systematic study requires new instruments with higher spectral resolutions and enough sensitivities.
In fact, this kind of assessment is one of the primary goals for planned X-ray instruments such as 
SXS (\cite{mitsuda10}) onboard \astroh (\cite{takahashi10}).
Using the SXS with an energy resolution better than 7~eV 
we could measure gas bulk motions in a fair number of X-ray bright clusters.
Furthermore, we may find line broadening originated from the gas turbulence, 
as a result of mergers, related shocks, or some activities in the massive black hole at cluster centers.
In addition, the SXS potentially will constrain
for the first time the line broadening from the thermal motion of ions (and hence the ion temperature). 
The present result proves that A~2256 is one of the prime targets for \astroh.
The expected spectra of the two components in A~2256 with the SXS
are shown in Fig.11 of \citet{takahashi10}.

\bigskip
We thank anonymous referee for useful comments.
We thank all the {\it Suzaku} team member for their supports.
To analyze the data, we used the ISAS Analysis Servers provided by ISAS/C-SODA.
KH acknowledges the support by a Grant-in-Aid for Scientific Research, No.21659292 from the MEXT.

% http://adsabs.harvard.edu/cgi-bin/nph-abs_connect?library&libname=a2256-references&libid=4ad8186ddd
% -> ads.tex
% grep cite intro.tex obs.tex ana.tex summary.tex >| cite.txt

\end{document}